\documentclass[prd,aps,preprintnumbers, showpacs, nofootinbib,superscriptaddress,
notitlepage
]{revtex4-1}
  \usepackage{latexsym,bm,amsmath,amssymb,amsfonts,color}
\newcommand{\beq}{\begin{eqnarray}}
\newcommand{\eeq}{\end{eqnarray}}

\newcommand{\nn}{\nonumber \\}

\newcommand{\Slash}[1]{{\ooalign{\hfil/\hfil\crcr$#1$}}}

\usepackage{graphicx}
\usepackage{amsmath,amsthm,amssymb}
\usepackage{color}

\begin{document}
\preprint{YITP-18-79}

\title{Holographic $J/\psi$ production near threshold  and the proton mass problem }

\author{Yoshitaka Hatta}
\affiliation{Yukawa Institute for Theoretical Physics, Kyoto University, Kyoto 606-8502, Japan}
\affiliation{Physics Department, Brookhaven National Laboratory, Upton, New York 11973, USA}

\author{Di-Lun Yang}
\affiliation{Nishina Center, RIKEN, Wako, Saitama 351-0198, Japan }


\begin{abstract}
It has been suggested that the production of a heavy quarkonium near threshold in electron-proton scattering can shed light on the origin of the proton mass via the QCD trace anomaly.   
We study the photoproduction  of $J/\psi$ off the proton using gauge/string duality and demonstrate that the $t$-dependence of the differential cross section $d\sigma/dt$ at small-$t$ is a sensitive probe of the trace anomaly. 
\end{abstract}
\maketitle

\section{Introduction}

From the early days of Quantum Chromodynamics (QCD), the origin of hadron masses has been a profound puzzle.  At the most naive level, one asks the question ``How can the QCD Lagrangian, written in terms of massless gluons and nearly massless quarks, give rise to the mass of the proton $M\sim 1$ GeV?''  More seriously, knowing that  
 energy and mass are equivalent in special relativity, one asks whether the `missing mass' comes from the relativistic orbital motion of quarks and gluons inside the proton. These kinetic energy contributions can be unambiguously defined and have been measured in deep inelastic scattering experiments \cite{Dulat:2015mca} as well as in lattice QCD simulations \cite{Hagler:2007xi,Alexandrou:2017oeh,Yang:2017erf}.  However, they  are not sufficient to account for the total mass. The fundamental reason the proton has a nonvanishing mass in the first place is because the approximate conformal symmetry of the classical QCD Lagrangian  is  broken by the quantum effects. This is quantified by the trace anomaly of the energy-momentum tensor $T^{\mu\nu}$
\beq
T^\mu_{\ \mu} = \frac{\beta(g)}{2g} F_a^{\mu\nu}F_{\mu\nu}^a+\cdots, \label{w}
\eeq
where $\beta(g)$ is the QCD beta function. The full decomposition formula thus reads  \cite{Ji:1994av}
\beq
M = M_q+M_g+M_m+M_a, \label{rule}
\eeq
where $M_{q/g}$ is the kinetic energy of quarks/gluons which comes from the traceless part of $T^{\mu\nu}$, $M_m$ is the current quark mass and $M_a \propto \langle P|T^{\mu}_{\mu}|P\rangle$ is the trace anomaly contribution. The decomposition  (\ref{rule})  is gauge invariant and well-defined, but is not entirely without controversy (see, e.g., \cite{Roberts:2016vyn,Lorce:2017xzd}). 

Recently, there has been a lot of interest among the nucleon structure community in determining the anomaly contribution $M_a$ \cite{workshop} as a key to understand the origin of the proton mass. Experiments dedicated to this goal have been proposed at the Jefferson Laboratory \cite{Joosten:2018gyo},   and the subject will likely continue to be discussed in the era of the future Electron-Ion Collider (EIC). Specifically, it has been proposed, based on some  theory suggestions \cite{Kharzeev:1995ij}, that one can access $M_a$ via the exclusive production of heavy quarkonium states such as $J/\psi$ at threshold in electron-proton scattering  $ep \to e'\gamma^* p\to  e'p'J/\psi$ \cite{Gittelman:1975ix,Camerini:1975cy,Binkley:1981kv,Kharzeev:1998bz,Brodsky:2000zc,Frankfurt:2002ka,Chekanov:2002xi,Gryniuk:2016mpk}. Heavy quarkonia are useful here because they only couple to gluons, not light quarks, and are therefore sensitive to the gluonic structure of the proton. However,  a  formula which relates the actual cross section to the trace anomaly has not been explicitly written down in the literature, although such a formula is crucial for the proper interpretation of the data. The main obstacle,  from the  perturbative QCD point of view, is that the QCD factorization for the twist-{\it four} operator $F^{\mu\nu}F_{\mu\nu}$ is difficult to establish despite the presence of a hard scale, the heavy quark mass. In view of this, one may seek alternative approaches which do not rely on the weak coupling/factorization framework. 

In this paper, we use the gauge/string duality to calculate the $J/\psi$ cross section in $ep$ collisions and study its connection to the trace anomaly.\footnote{Vector meson production at high energy  has been previously studied in holographic frameworks \cite{Forshaw:2012im,Costa:2013uia,Lee:2018zud}, but not in connection with the proton mass problem.} 
This approach allows us to bypass the issue of factorization and directly evaluate the scattering amplitude in string/gravity theory dual to QCD (or QCD-like theories). The original version of the duality is for conformal theories in which particles are massless and $T^{\mu\nu}$ is traceless. Subsequently,  it has been generalized to theories with conformal symmetry breaking so that the problem of the proton mass can  be addressed. 
Our work is distinct from the previous works on the application of gauge/string duality to high energy, lepton-hadron deep inelastic scattering (see e.g., \cite{Polchinski:2002jw,Brower:2006ea,Hatta:2007he,BallonBayona:2008zi,Hatta:2009ra,Cornalba:2009ax,BallonBayona:2010ae,Koile:2011aa,Costa:2012fw,Nishio:2014rya,Watanabe:2015mia,Kovensky:2018xxa}) where scattering amplitudes are dominated by the exchange of the graviton. Because of its spin-2 nature, the graviton exchange predicts a too steep rise of cross sections with increasing energy to be compatible with the experimental data.\footnote{This problem may be cured by modifying the graviton or adding unitarity corrections. We do not pursue these  directions in the present paper.} Here instead, we apply gauge/string duality to low energy scattering where the relevant momentum scales are on the order of a few GeV. Near the threshold, the cross section rises from zero, and our idea is to explain this behavior by the graviton exchange picture. An interesting complication in this regime is that  the contribution from other supergravity modes can become equally important. In particular, we shall be interested in the exchange of the {\it dilaton} which, according to the AdS/CFT correspondence, is dual  to the operator $F^{\mu\nu}F_{\mu\nu}$ in (\ref{w}).

   We work in the simplest setup to introduce heavy quarks (the so-called `D3/D7 model' \cite{Kruczenski:2003be}) and compute the cross section of the subprocess $\gamma p \to p'J/\psi$ in the photoproduction limit. We consider both the graviton and dilaton exchanges in an asymptotically AdS space, and relate this amplitude to the matrix elements of the traceless and trace parts of the energy momentum tensor. Our goal is to write down a formula for the differential cross section $d\sigma/dt$ which explicitly depends on the gluon condensate $\langle P|F^{\mu\nu}F_{\mu\nu}|P\rangle$, and quantitatively study its impact on the shape of the $t$-distribution.

This paper is structured as follows. In Section II, we give a brief review of the nucleon mass sum rule (\ref{rule}) and discuss the  nonforward matrix element of the QCD energy momentum tensor. In Section III, we explain the basic kinematics of the $\gamma p \to J/\psi p'$ process. In Section IV, we compute the cross section by using gauge/string duality and numerically evaluate the differential cross section $d\sigma/dt$.   We then conclude in Section V. 

\section{Nucleon mass and the QCD energy momentum tensor}

\subsection{Nucleon mass decomposition}

We begin by  briefly reviewing how  the formula (\ref{rule}) is derived from the  QCD energy momentum tensor.  
Consider  the standard matrix elements
\beq
\langle P|T^{\mu\nu}|P\rangle=2P^\mu P^\nu, \qquad \langle P|T^\alpha_{\ \alpha}|P\rangle = 2M^2,
\eeq
where the proton single particle state is normalized as $\langle P'|P\rangle=2P^0(2\pi)^3\delta^{(3)}(\vec{P}-\vec{P}')$. 
We write the energy momentum tensor in the form 
\beq
T^{\mu\nu} = 
-F_a^{\mu\lambda}F^{a\nu}_{\ \lambda} +\frac{\eta^{\mu\nu}}{4}F_a^{\alpha\beta}F^a_{\alpha\beta}+ i\bar{\psi} \gamma^{(\mu} D^{\nu)} \psi 
=T_g^{\mu\nu} + T_q^{\mu\nu}, \label{tmu}
\eeq 
 where $\eta^{\mu\nu}=(1,-1,-1,-1)$.\footnote{We shall use this `mostly minus' metric throughout this paper, differently from most of the literature on gauge/string duality.} Throughout this paper, we use the notation $A^{(\mu}B^{\nu)}\equiv \frac{A^\mu B^\nu+A^\nu B^\mu}{2}$. As is well known, the trace of this energy momentum tensor contains the QCD trace anomaly. 
\beq
T^\alpha_{\ \alpha} = \frac{\beta(g)}{2g}F_a^{\alpha\beta}F^a_{\alpha\beta} + m(1+\gamma_m)\bar{\psi}\psi, \label{trace}
\eeq
where $\beta(g)= -\frac{g^3}{16\pi^2}\left(\frac{11N_c}{3}-\frac{2n_f}{3}\right)+\cdots$ is the QCD beta function, $m$ is the current quark mass, and $\gamma_m$ is the anomalous dimension of the mass operator.  
One can decompose the tensor into the traceless and trace parts (in $d=4$ dimensions)
\beq
T^{\mu\nu} = \left( T^{\mu\nu}-\frac{\eta^{\mu\nu}}{d}T^{\alpha}_{\ \alpha}\right) +\frac{\eta^{\mu\nu}}{d}T^{\alpha}_{\ \alpha} \equiv \bar{T}^{\mu\nu} +\hat{T}^{\mu\nu}.
\eeq
The traceless part $\bar{T}^{\mu\nu}$ can be further decomposed into the quark and gluon parts $\bar{T}^{\mu\nu}=T_{q,kin}^{\mu\nu} + T_{g,kin}^{\mu\nu}$ which can be interpreted as the kinetic energy contributions. Also, the trace part (\ref{trace}) is a sum of the mass and anomaly contributions $\hat{T}^{\mu\nu} = T_m^{\mu\nu} + T_a^{\mu\nu}$. We can thus write 
\beq
T^{\mu\nu}=  T_{q,kin}^{\mu\nu}+T_{g,kin}^{\mu\nu} +T_m^{\mu\nu} + T_a^{\mu\nu}.
\eeq
From Lorentz symmetry, their matrix elements can be parameterized as 
\beq
\langle P|T^{\mu\nu}_{q,kin}|P\rangle &=& 2a(\mu^2)\left(P^\mu P^\nu -\frac{\eta^{\mu\nu}}{4}M^2\right), \label{tra1}
\\
\langle P|T^{\mu\nu}_{g,kin}|P\rangle &=& 2(1-a(\mu^2))\left(P^\mu P^\nu -\frac{\eta^{\mu\nu}}{4}M^2\right),
\\
\langle P|T^{\mu\nu}_m|P\rangle &=& \frac{1}{2}b(\mu^2)\eta^{\mu\nu}M^2,\\
\langle P|T^{\mu\nu}_a|P\rangle &=& \frac{1}{2}(1-b(\mu^2))\eta^{\mu\nu}M^2, \label{tra4}
\eeq
where $\mu^2$ is the renormalization scale.
Let us now work in the rest frame and define the Hamiltonian $H_i=\int d^3x T^{00}_i$. We can then write 
\beq
M=M_q+M_g+M_m + M_a, \label{3}
\eeq
where 
\beq
M_q&=& \frac{\langle P|H_q|P\rangle}{\langle P|P\rangle} = \frac{3a}{4}M \\
M_g &=& \frac{\langle P|H_g|P\rangle}{\langle P|P\rangle} =\frac{3(1-a)}{4}M\\
M_m &=& \frac{\langle P|H_m|P\rangle}{\langle P|P\rangle} =\frac{b}{4}M\\
M_a &=& \frac{\langle P|H_a|P\rangle}{\langle P|P\rangle} =\frac{1-b}{4}M  \label{anomal}
\eeq 
(Note that $\langle P|P\rangle = 2M\int d^3x$.) 
We see that the trace part $M_m+M_a$ accounts for a quarter of the proton mass. 
Ji proposed a slightly different decomposition   \cite{Ji:1994av} by reshuffling terms in   (\ref{3}). From the equation of motion, one can write  
\beq
T^{00}_{q,kin}= i\bar{\psi}D^0\gamma^0\psi -\frac{m}{4}\bar{\psi}\psi +\cdots= i\bar{\psi}\vec{D}\cdot \vec{\gamma}\psi+ \frac{3m}{4}\bar{\psi}\psi + \cdots.
\eeq
It is more reasonable to interpret the last term as a part of the quark mass contribution. By moving this term into $T^{00}_m$, one gets an alternative decomposition
\beq
M=\widetilde{M}_q+\widetilde{M}_g+\widetilde{M}_m + \widetilde{M}_g 
\eeq
 where 
\beq
\widetilde{M}_q&=& \frac{\langle P|H_q|P\rangle}{\langle P|P\rangle} = \frac{3}{4}\left(a-\frac{b}{1+\gamma_m}\right)M, \\ 
\widetilde{M}_g &=& \frac{\langle P|H_g|P\rangle}{\langle P|P\rangle} =\frac{3(1-a)}{4}M,\\
\widetilde{M}_m &=& \frac{\langle P|H_m|P\rangle}{\langle P|P\rangle} =\frac{b}{4} \frac{4+\gamma_m}{1+\gamma_m}M, \\ 
\widetilde{M}_a &=& \frac{\langle P|H_a|P\rangle}{\langle P|P\rangle} =\frac{1-b}{4}M. 
\eeq 
The parameter $a(\mu^2)$ is related to the matrix element of the quark and gluon twist-two operators, and can be extracted from the experimental data of deep inelastic scattering. It is more difficult to access the parameter $b(\mu^2)$.  Being associated with the twist-four operator $F^2$, any dependence on $b$ is strongly suppressed in high energy scattering.  Instead, one should look at low-energy scattering.

\subsection{Non-forward proton matrix element}

In the actual experimental process $ep\to e'p'J/\psi$, one cannot directly access the forward matrix element $\langle P|T^{\mu\nu}|P\rangle$ because it is kinematically impossible. In practice, experimentalists measure the non-forward matrix element $\langle P'|T^{\mu\nu}|P\rangle$ and extrapolate it to the forward limit 
$\Delta^\mu=P'^\mu -P^\mu\to 0$.  
The general parameterization of the non-forward matrix element of $T_{q,g}^{\mu\nu}$  for a spin-$\frac{1}{2}$ hadron is \cite{Ji:1996ek}
\beq
\langle P'|T^{\mu\nu}_{q,g}|P\rangle = \bar{u}(P')\Bigl[ A_{q,g}\gamma^{(\mu}\bar{P}^{\nu)} + B_{q,g}\frac{\bar{P}^{(\mu}i\sigma^{\nu)\alpha}\Delta_\alpha}{2M} + C_{q,g}\frac{\Delta^\mu\Delta^\nu-g^{\mu\nu}\Delta^2}{M} + \bar{C}_{q,g}M\eta^{\mu\nu} \Bigr] u(P) \nn 
= \bar{u}(P')\Bigl[ (A_{q,g}+B_{q,g})\gamma^{(\mu}\bar{P}^{\nu)} -\frac{\bar{P}^\mu \bar{P}^\nu}{M}B_{q,g} + C_{q,g}\frac{\Delta^\mu\Delta^\nu-g^{\mu\nu}\Delta^2}{M} + \bar{C}_{q,g}M\eta^{\mu\nu} \Bigr] u(P), \label{jid}
\eeq
 where $\bar{P}^\mu\equiv \frac{P^\mu +P'^\mu}{2}$. 
 In the second line we used the Gordon identity. $A,B,C,\bar{C}$ all depend on $\Delta^2=t$ (and also on the renormalization scale). In the literature, often the notation $D_{q,g}(t)=4C_{q,g}(t)$ is used, and is called the `D-term'. 
Multiplying both sides by $\partial^\mu \sim \Delta^\mu$, we see that all terms on the right hand side except the $\bar{C}_{q,g}$ term vanish  $\langle \partial_\mu T^{\mu\nu}_{q,g} \rangle \sim \Delta^\nu \bar{C}_{q,g}$.  Since the sum $T^{\mu\nu}_q + T^{\mu\nu}_g$ is conserved, $\bar{C}_q+\bar{C}_g=0$. 

Taking the trace of (\ref{jid}) we find\footnote{In dimensional regularization, the anomaly entirely comes from the gluon part $T_g$, see, e.g., \cite{Nielsen:1977sy}.}
\beq
\langle P'|(T_{g})^\mu_\mu|P\rangle& =& \langle P'|\left(\frac{\beta(g)}{2g}F^a_{\mu\nu}F^{\mu\nu}_a+m\gamma_m \bar{\psi}\psi\right)|P\rangle \nn  
&=& \bar{u}(P')\Bigl[ A_{g} M + \frac{B_{g}}{4M}\Delta^2 -3\frac{\Delta^2}{M}C_{g} +4 \bar{C}_{g}M \Bigr] u(P).   \label{anomaly}
\eeq
From this we can deduce that
\beq
&& \langle P'|\frac{\beta(g)}{2g}F^a_{\mu\nu}F^{\mu\nu}_a|P\rangle  \nn
&& = \bar{u}(P')\Bigl[ (A_{g} -\gamma_mA_q) M + (B_g-\gamma_m B_q)\frac{\Delta^2}{4M} -3\frac{\Delta^2}{M}(C_{g}-\gamma_m C_q) +4 (\bar{C}_{g}-\gamma_m \bar{C}_q)M \Bigr] u(P).   \label{anomaly}
\eeq
Comparing with (\ref{tra1})--(\ref{tra4}), we find the following relations 
\beq
 A_q(0)=a, &\quad& A_g(0)=1-a,  
 \eeq
 and 
 \beq
b=(A_q(0)+4\bar{C}_q(0) )(1+\gamma_m), &\quad&
 1-b = (A_g(0)+ 4\bar{C}_g(0)) (1+ \gamma_m) -\gamma_m .
\eeq

For a later purpose, let us define the `transverse-traceless'  part of $T_g^{\mu\nu}$. First consider the transverse part of  $T_g^{\mu\nu}$  
\beq
T_{g\perp}^{\mu\nu}\equiv T_g^{\mu\nu}-\frac{1}{\square}\partial^\mu \partial_{\alpha}T_g^{\nu\alpha}-\frac{1}{\square}\partial^\nu \partial_{\alpha}T_g^{\mu\alpha}+\frac{1}{\square^2}\partial^\mu \partial^\nu \partial_{\alpha}\partial_\beta T_g^{\alpha\beta},
\eeq
where $\square=\partial^{\mu}\partial_{\mu}$, such that $\partial_\mu T_{g\perp}^{\mu\nu}=0$. 
Its matrix element  can be readily inferred from (\ref{jid})
\beq
\langle P'|T_{g\perp}^{\mu\nu}|P\rangle &=& \langle P'| \left(T_g^{\mu\nu}-\frac{1}{\Delta^2}\Delta^\mu \Delta_{\alpha}T_g^{\nu\alpha}-\frac{1}{\Delta^2}\Delta^\nu \Delta_{\alpha}T_g^{\mu\alpha}+\frac{1}{\Delta^4}\Delta^\mu \Delta^\nu \Delta_{\alpha}\Delta_\beta T_g^{\alpha\beta} \right)|P\rangle 
\nn 
&=& \bar{u}(P')\Biggl[ (A_{g}+B_{g})\gamma^{(\mu}\bar{P}^{\nu)} -\frac{\bar{P}^\mu \bar{P}^\nu}{M}B_{g} +\left( \frac{\Delta^2}{M} C_{g} -\bar{C}_g M\right)  \left(\frac{\Delta^\mu\Delta^\nu}{\Delta^2} -\eta^{\mu\nu}\right) \Biggr] u(P). \nonumber
\eeq
We then define the transverse-traceless (TT) part by making $T_\perp^{\mu\nu}$ traceless while preserving its transverse property 
\beq
T_{gTT}^{\mu\nu} \equiv T_{g\perp}^{\mu\nu} +\frac{1}{3}\left(\frac{\partial^\mu\partial^\nu}{\square}-\eta^{\mu\nu}\right)T_{g\perp \alpha}^\alpha. 
\eeq
This has the following matrix element
\beq
\langle P'|T_{gTT}^{\mu\nu}|P\rangle = \bar{u}(P')\Biggl[ (A_{g}+B_{g})\gamma^{(\mu}\bar{P}^{\nu)} -\frac{\bar{P}^\mu \bar{P}^\nu}{M}B_{g}   +\frac{1}{3} \left(\frac{\Delta^\mu\Delta^\nu}{\Delta^2}-\eta^{\mu\nu}\right)\left(A_g M +\frac{\Delta^2}{4M}B_g\right) \Biggr]u(P).
\label{tt} \nn 
\eeq
Note that the forward limit is ambiguous as it depends on the angle of $\vec{\Delta}$.
\beq
\lim_{P'\to P} \langle   P'|T_{gTT}^{\mu\nu}|P\rangle = \lim_{\Delta \to 0} 2A_g \left(P^\mu P^\nu + \frac{M^2}{3}\left(\frac{\Delta^\mu \Delta^\nu}{\Delta^2}-\eta^{\mu\nu}\right) \right).
\eeq

\section{Exclusive photoproduction of $J/\psi$ in $ep$ scattering }

In this section we briefly review the basic kinematics of the process $ep\to e'\gamma^*p \to e'p' J/\psi $ which will be studied at the Jefferson Laboratory and possibly at the future EIC  
\cite{Joosten:2018gyo}. The connection to the trace anomaly will be discussed in the next section. 
The electron part can be factored out, so in practice one considers the subprocess $\gamma^*(q)p(P) \to p(P') J/\psi(k) $.  The cross section is given by the formula  
\beq
\sigma(\gamma p\to pJ/\psi) &=&\frac{e^2}{4MK} \int \frac{d^3k}{2E_k(2\pi)^3}\frac{d^3P'}{2E'(2\pi)^3} (2\pi)^4\delta^{(4)}(P+q-P'-k)\langle P|\epsilon \cdot J(0) |P'k\rangle \langle P'k|\epsilon^*\cdot J(0)|P\rangle \nn 
&=&\frac{e^2 k_{cm}}{64\pi^2 MKW } \int d\Omega \langle P|\epsilon \cdot J |P'k\rangle \langle P'k|\epsilon^*\cdot J|P\rangle 
\eeq 
where $K=\frac{2P\cdot q-Q^2}{2M} = \frac{W^2-M^2}{2M}$, and $W^2=(P+q)^2$ is the virtual photon-proton center-of-mass (COM) energy $(Q^2=-q^2)$.   Since the integral is Lorentz invariant, it can be conveniently evaluated in the photon-proton COM frame, which was done in the second line. We also defined 
\beq
k_{cm}^2= \frac{(W^2-(M_\psi +M)^2)(W^2-(M_\psi -M)^2)}{4W^2},
\eeq
as  the $J/\psi$ momentum in the COM frame. ($M_\psi$ denotes the mass of $J/\psi$.) Switching back to the Lorentz invariant variable  $t=(P-P')^2 = 2M^2-2(EE'-|P||p|\cos \theta)$ we get 
\beq
\sigma(\gamma p\to p'J/\psi)= \frac{e^2}{64\pi MKW|P_{cm}| } \int dt  \langle P|\epsilon \cdot J |P'k\rangle \langle P'k|\epsilon^*\cdot J|P\rangle 
\eeq
 where
\beq
|P_{cm}|^2= \frac{W^4-2W^2(M^2-Q^2)+(M^2+Q^2)^2}{4W^2}
\eeq
is the incoming proton momentum in the COM frame.  In the 
photoproduction limit $q^2=-Q^2\to 0$,  only the transverse polarizations survive and we find
\beq
\sigma(\gamma p\to p'J/\psi)= \frac{e^2}{16\pi (W^2-M^2)^2 }  \frac{1}{2}\sum_i^{1,2} \int dt  \langle P|\epsilon_i \cdot J(0) |P'k\rangle \langle P'k|\epsilon_i^*\cdot J(0)|P\rangle . \label{cro}
\eeq

The $t$-integral in (\ref{cro}) is for $0>t_{min}>t>t_{max}$. Ideally, one would like to study  the forward matrix element $t=0$, but this is kinematically not allowed. In practice one has to extrapolate the amplitude from $t\lesssim t_{min}$ to $t\to 0$.

\begin{figure}[h]
 \includegraphics[height=61mm]{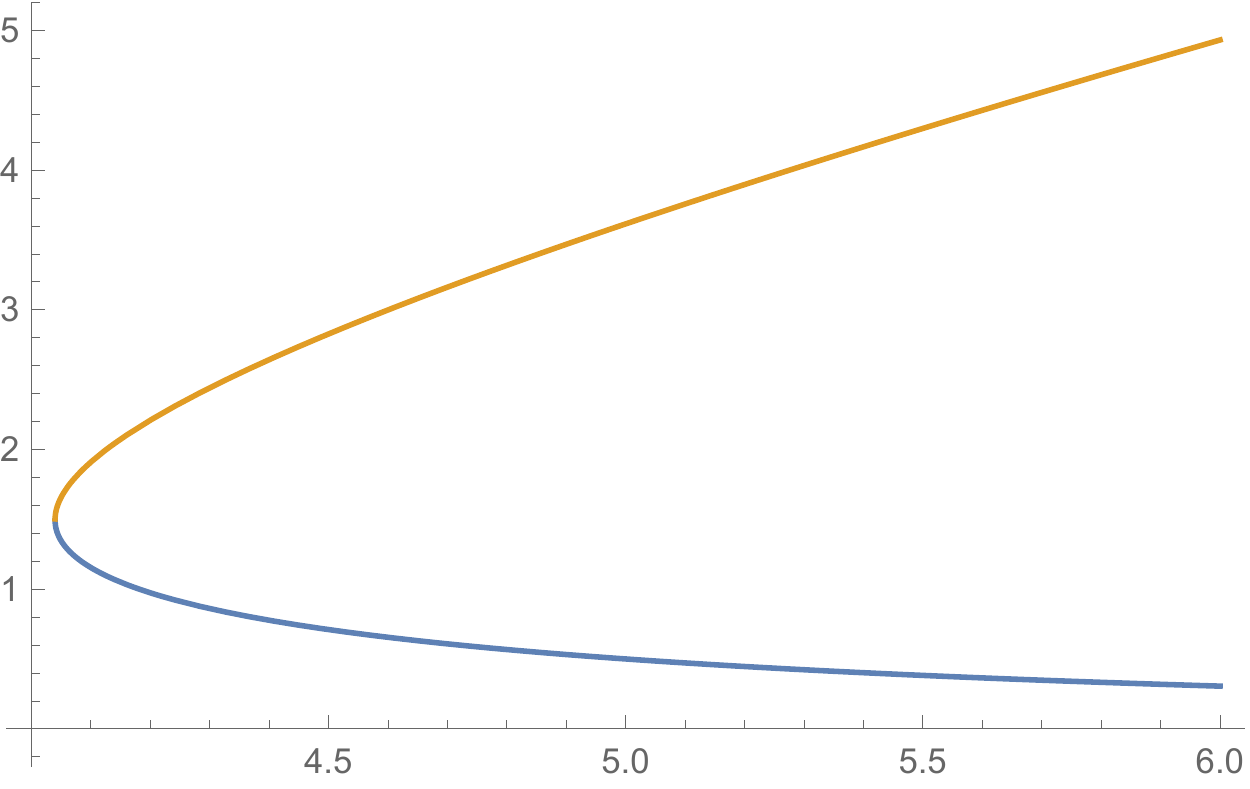} 
\caption{ $|\Delta_{max}|=\sqrt{-t_{max}}$ (upper curve) and  $|\Delta_{min}|=\sqrt{-t_{min}}$ (lower curve) in units of GeV as a function of $W>W_{th}$ with  $M=0.94$ GeV and $M_\psi=3.1$ GeV. 
} 
\label{fig1}
\end{figure}

To find $t_{min}$ we again work in the COM frame and take the photoproduction limit $Q^2\to 0$. Then $P_{cm}=\frac{W^2-m^2}{2W}$ and 
\beq
t&=& (\sqrt{P_{cm}^2+M^2}-\sqrt{k_{cm}^2+M^2})^2-(\vec{P}_{cm}+\vec{k}_{cm})^2  \nn
&\le&
 (\sqrt{P_{cm}^2+M^2}-\sqrt{k_{cm}^2+M^2})^2-(|P_{cm}|-|k_{cm}|)^2 \equiv t_{min}  \label{min}
\eeq
This gives a complicated function of $W$. At the threshold $W=W_{th}=M+M_\psi\approx 4.04$ GeV, we get 
\beq
t_{min}=  -\frac{MM_{\psi}^2}{M+M_\psi} \approx - (1.5\ {\rm GeV})^2.
\eeq
   where we used $M\approx 0.94$ GeV and $M_\psi\approx 3.10$ GeV.
At large $W$, on the other hand, we find
\beq
t_{min} =- \frac{M^2M_\psi^4}{W^4}+\cdots, \qquad t_{max} = -W^2 + \cdots
\eeq
In Fig.~\ref{fig1}, we plot  $|\Delta_{min}|=\sqrt{-t_{min}}$ and  $|\Delta_{max}|=\sqrt{-t_{max}}$ as a function of $W>W_{th}$. Away from the threshold, $\Delta_{min}$ decreases rapidly and becomes negligible compared to the other mass scales.

We shall be interested in the `threshold region' which we loosely define as $W_{th} \le W \lesssim 6$ GeV, or in terms of the photon energy in the proton rest frame, 
\beq
\frac{W^2_{th}-M^2}{2M} \approx 8.2\ {\rm GeV} \le E_\gamma \lesssim 20 \ {\rm GeV}.
\eeq
While the considered energy range is rather narrow, it is actually sufficient to discuss the effect of the trace anomaly, as we shall demonstrate in the following.

\section{Holographic computation of the cross section} 

\subsection{Setup}
In the cross section formula (\ref{cro}), the difficult part is the nonperturbative matrix element 
\beq
 \langle P|\epsilon_i(q) \cdot J(q) |P'k\rangle =(2\pi)^4\delta^{(4)}(P+q-P'-k) \langle P|\epsilon_i(q) \cdot J(0) |P'k\rangle.
\eeq
 In this section we evaluate this using holography. Our setup is as follows. The four-dimensional Minkowski space is located at the boundary of a five-dimensional, asymptotically anti-de Sitter (AdS) space with the metric
\beq
ds^2 =g_{MN}dx^M dx^N \approx R^2\frac{\eta^{\mu\nu}dx_\mu dx_\nu -dz^2}{z^2},  \qquad (z\approx 0) 
\eeq
where $R$ is the AdS radius.  $z$ denotes the fifth dimension and the boundary is  at $z=0$. 
In the infrared region (large $z$), the metric is modified such that the dual theory  breaks conformal symmetry and contains light quarks/hadrons. The precise way in which these modifications are done is not important for our purpose. We simply assume that the theory contains baryons  which are described by an unspecified bulk  action  $S_B$. (We have in mind models such as, for example, in \cite{Callan:1999zf,Abidin:2009hr,Avsar:2009hc}.)  

We introduce `charm' quarks in the theory by adding one D7 brane with the action 
\beq 
S_{D7}&=& - T_{D7} \int d^8\bar{\xi} e^{-\phi}\sqrt{-{\rm det}(G_{ab}+2\pi \alpha' {\cal F}_{ab})} \nn 
&=&  - T_{D7}\int d^8\bar{\xi} e^{-\phi} \sqrt{-G}\left(1+\frac{(2\pi \alpha')^2}{4} {\cal F}_{ab}{\cal F}^{ab}+\cdots\right), \label{d7}
\eeq
where $T_{D7}=(32\pi^6g^2\alpha'^4)^{-1}$ denotes the D7-brane tension, $\phi$ is the dilaton and $G_{ab}=g_{\bar{M}\bar{N}}\partial_{\bar{\xi}^a}x^{\bar{M}} \partial_{\bar{\xi}^b}x^{\bar{N}}$ is the induced metric. In addition, $\bar{\xi}^{a(b)}$ denote the world-volume coordinates on the D7 branes, while $x^{\bar{M}(\bar{N})}$ represent the ten-dimensional spacetime coordinates in $\text{AdS}_5\times\text{S}^5$.
 The worldvolume of the D7 brane wraps  $S^3 \in S^5$ and extends in the fifth dimension from $z=0$ to $z=z_m$ where $z_m$ is inversely proportional to the heavy quark mass $m_q$ as $z_m=\frac{\sqrt{g^2N_c}}{2\pi m_q}$. Explicitly, the induced metric reads
\beq
ds^2_{D7}= \frac{R^2}{z^2} \eta_{\mu\nu}dx^\mu dx^\nu -\frac{R^2}{z^2\left(1-\frac{z^2}{z_m^2}\right)}dz^2 -\left(1-\frac{z^2}{z_m^2}\right) R^2 d\Omega_3^2\,.  \label{ind}
\eeq
 An important point to emphasize is that the supports of $S_B$ and $S_{D7}$ are well separated in the $z$ direction: $z_m$ is much smaller than the typical $z$ values of the baryon wavefunction. The latter is a normalizable mode localized around $z\sim 1/\Lambda_{QCD}$. 

In (\ref{d7}), ${\cal F}$ represents the field strength coming from  gauge field fluctuations. It can be decomposed into two parts, 
\beq
{\cal F}=\bar{F}+ F,
\eeq
 where  $\bar{A}^\mu$ and $A^\mu$ correspond to heavy  vector mesons (such as $J/\psi$) and the electromagnetic gauge field (photon), respectively. The wavefunction of an on-shell photon with momentum $q^\mu$ ($q^2=0$) is simply a plane wave 
\beq
A_\mu \propto \epsilon_\mu e^{iq\cdot x}. \label{amu}
\eeq 
where $\epsilon^\mu(q)$ is the polarization vector with the property $\epsilon \cdot q=0$. 
The spectrum of vector mesons is well understood in this model \cite{Kruczenski:2003be}. They are characterized by the normalizable bulk wavefunction 
\beq
\bar{A}_\mu \propto \xi_\mu \phi_{n,l}(z) e^{-ik\cdot x}Y^l(S^3), \label{fun}
\eeq
 and their masses are given by
\beq
M_{n,l}= \frac{2\sqrt{(n+l+1)(n+l+2)}}{z_m}.
\eeq
 $\xi^\mu$ is the vector meson polarization vector which satisfies  $\xi(k)\cdot k=0$, and $Y^l$ is the spherical harmonics on $S^3$.
We may identify the lightest state $n=l=0$ with $J/\psi$. It has mass $M_\psi=\frac{2\sqrt{2}}{z_m} = \frac{4\sqrt{2}\pi m_q}{\sqrt{g^2N_c}}$ and wavefunction
\beq
\phi_{n=l=0} = \frac{z^2}{z_m^2}, \label{wav}
\eeq

We have not specified the proportionality constant in (\ref{amu}) and (\ref{fun}) (see, however, \cite{Hong:2003jm,BallonBayona:2008zi,Nishio:2014rya}).  Fixing this amounts to fixing the strength of the coupling ${\cal F}^2 \sim \bar{F}F$ between the photon and $J/\psi$, and hence the overall normalization of the cross section. Instead of introducing extra assumptions,  we treat the overall factor as a free parameter to be  fitted to the experimental data. Our prediction, then, is the $t$-dependence of the differential cross section $d\sigma/dt$. As we demonstrate in what follows, the shape of $d\sigma/dt$ is sensitive to the QCD trace anomaly.   


\subsection{Scattering amplitude}

We now explain how we evaluate the matrix element $\langle P|\epsilon\cdot J |P'k\rangle$. In the framework of gauge/string duality, the current insertion $J(q)$ on the boundary field theory creates a gauge field excitation in the bulk AdS space. This scatters off the bulk proton field  via graviton and dilaton exchanges. This amplitude, the so-called Witten diagram, is evaluated as (see e.g., \cite{Polchinski:2002jw,BallonBayona:2008zi,Abidin:2009hr,Hatta:2010kt})
\beq
\langle P| \epsilon\cdot J(q)|P'k\rangle  &=&\frac{i}{f_\psi}\int d^4xdz e^{i(q-k)\cdot x} \int d^4x'dz' e^{i(P-P')\cdot x'} \label{mat} \\ && \times 
 \Biggl( \frac{\delta S_{D7}(q,k,z)}{\delta g_{MN}}  G_{MNM'N'}(xz,x'z') \frac{\delta S_{B}(P,P',z)}{\delta g_{M'N'}} 
+\frac{\delta S_{D7}}{\delta \phi(xz)}  D(xz,x'z') \frac{\delta S_{B}}{\delta \phi(x'z')} \Biggr),  \nonumber
\eeq
where  $G_{MNM'N'}$ and $D$ are the graviton and dilaton bulk-to-bulk propagators, respectively. 
  $f_\psi$ is the decay constant  defined as 
 $\langle 0|{\cal O}_\psi^\mu(0)|k\rangle = f_\psi \xi^\mu$, where  ${\cal O}^\mu_\psi$ is an interpolating operator of  $J/\psi$. The notation $\delta S/\delta g_{MN}$ (or $\delta S/\delta \phi$) means that after the coupling to the graviton (or dilaton) is extracted, the action is evaluated with the on-shell bulk wavefunctions of the external states (including the polarization part $\epsilon^\mu,\xi^\mu,\bar{u}(P),u(P')$). The plane-wave phases are trivial and have been factored out in (\ref{mat}). 

Since $\delta S/\delta g_{MN}$ is covariantly conserved, the amplitude is gauge (diffeomorphism) invariant and can be evaluated in any convenient gauge. For our purpose, it is crucial to work in the transverse-traceless (TT) gauge \cite{Garriga:1999yh}
\beq
\delta g_{Mz}=0, \qquad \delta g^\mu_\mu = \nabla_\mu \delta g^{\mu\nu}=0,
\eeq
for the metric fluctuations 
\beq
g_{MN}=g^{AdS}_{MN} + \delta g_{MN}.
\eeq
In this gauge, only the $MN=\mu\nu$ components in (\ref{mat}) survive. Moreover, all the components  $\delta g_{\mu\nu}$ are decoupled in the equation of motion. We argue that  in this gauge one can make a connection between (\ref{mat}) and the matrix element of $T^{\mu\nu}$. (See Ref.~\cite{Abidin:2008hn} for a related discussion.) To see this, note that the $z$ integral in (\ref{mat}) is restricted to a small region $z<z_m$ near the boundary. In this region, the bulk-to-bulk propagators $G_{MNM'N'}$ and $D$ essentially become the boundary-to-bulk propagators up to a proportionality constant $\propto z^4$. The latter are associated with the so-called non-normalizable modes which are excited by the insertion of  dual boundary operators $T^{\mu\nu}$ and $F^{\mu\nu}F_{\mu\nu}$, respectively. In the graviton sector, such a direct connection  is most transparent in the TT gauge where the $M,N=z$ components of the propagator are eliminated. The details of this `matching' is presented in Appendix \ref{a1}. Based on this, we  rewrite  (\ref{mat}) as
\beq
 \langle P| \epsilon\cdot J(0)|P'k\rangle &\approx &-\frac{2\kappa^2}{f_\psi R^3}\int^{z_m}_0 dz \frac{\delta S_{D7}(q,k,z)}{\delta g_{\mu\nu}} \frac{z^2R^2}{4} \langle P|T_{\mu\nu}^{gTT}|P'\rangle \nn 
&&  + \frac{2\kappa^2}{f_\psi R^3 } \frac{3}{8}  \int^{z_m}_0  dz \frac{\delta S_{D7}(q,k,z)}{\delta \phi}  \frac{z^4}{4} \langle P| \frac{1}{4}F^{\mu\nu}_a F^{a}_{\mu\nu} |P'\rangle,  \label{master}
\eeq
where $2\kappa^2=\frac{8\pi^2}{N_c^2}R^3$ is the five-dimensional gravitational constant.  $T^{\mu\nu}_{gTT}$ is the transverse-traceless part of the gluon energy momentum tensor introduced in the previous section. We have removed the momentum-conserving delta function $(2\pi)^4\delta(P+q-P'-k)$. Note that only the gluon part of $T^{\mu\nu}$ appears. This is because the graviton has been emitted by a $J/\psi$, and we know in QCD that heavy quarkonia only couple to gluons, not light quarks. Holographically, this is manifested by the fact that the quarkonium-graviton coupling occurs in the asymptotically AdS region $z\sim 0$ where the theory is dual to pure gluodynamics and heavy quarks, while the light quark degrees of freedom reside at much larger values of $z$.\footnote{Note also that $T_q^{\mu\nu}\sim {\cal O}(N_c)$ is subleading compared to $T_g^{\mu\nu}\sim {\cal O}(N_c^2)$ in the large-$N_c$ limit.}  

It is important to mention that after the approximation mentioned above, the amplitude (\ref{master}) now depends on the matrix element of {\it local} gluonic operators. This is indeed what one expects in the low energy, near-threshold region \cite{Kharzeev:1998bz,Brodsky:2000zc}.   At high energy, on the other hand, the amptitude is sensitive to nonlocal gluonic operators and  the corresponding generalized parton distributions.

\subsection{Graviton and dilaton couplings}

Next we proceed to compute the graviton and dilaton couplings to the external states. It is straightforward to evaluate the photon-vector meson-graviton coupling $\delta S_{D7}/\delta g_{\mu\nu}$. Similarly to \cite{Hatta:2010kt}, we find
\beq
\delta S_{D7}= -K_{D7}\int d\Omega_3^2Y^l(S^3) \int d^4xdz \frac{R^5}{z^5}\left(1-\frac{z^2}{z_m^2}\right) 
\Biggl[ \left(F^{\mu\rho}\bar{F}^\nu_{\ \rho} + \bar{F}^{\mu\rho}F^\nu_{\ \rho} -\frac{\eta^{\mu\nu}}{2}F_{\alpha\beta}\bar{F}^{\alpha\beta}\right)\delta g_{\mu\nu} \nn  -\frac{g^{zz}}{2}F_{\alpha\beta}\bar{F}^{\alpha\beta}\delta g_{zz} + 2\bar{F}^{z\rho }F^\mu_{\ \rho} \delta g_{z\mu} \Biggr],
\eeq
where $K_{D7}\equiv \frac{ N_f T_{D7}(2\pi\alpha')^2}{2}R^3$ and we used $F^{z\mu}=0$. Note that $F^{\mu\rho}\bar{F}^\nu_{\ \rho} = G^{\mu\alpha}G^{\rho\beta}F_{\alpha\beta}G^{\nu\lambda}\bar{F}_{\lambda\rho} = (z/R)^6\eta^{\mu\alpha}\eta^{\rho\beta}F_{\alpha\beta}\eta^{\nu\lambda}\bar{F}_{\lambda\rho}$.  In the TT gauge,  we only have to consider $\delta g_{\mu\nu}$ and find 
\beq
\frac{\delta S_{D7}}{\delta g_{\mu\nu}} \propto  -2K_{D7}\int d\Omega_3^2Y^l(S^3)\frac{z}{R}\phi(z) \left(1-\frac{z^2}{z_m^2}\right)  \left(\Pi^{\mu\nu}-\frac{\eta_{\alpha\beta}\Pi^{\alpha\beta}}{4}\eta^{\mu\nu}\right), \label{last}
\eeq
where 
\beq
\Pi^{\mu\nu}(q,k) \equiv q^{(\mu} k^{\nu)} \epsilon\cdot \xi + \epsilon^{(\mu} \xi^{\nu)} q\cdot k-q^{(\mu} \xi^{\nu)}  k\cdot \epsilon -k^{(\mu} \epsilon^{\nu)} q\cdot \xi.
\eeq
The proportionality symbol in (\ref{last}) is because of the normalization issue mentioned below (\ref{wav}). 
The term proportional to $\eta^{\mu\nu}$ in (\ref{last}) drops out when contracted with the traceless tensor $T^{gTT}_{\mu\nu}$ in (\ref{master}). 
 
On the other hand, computing the photon-vector meson-dilaton coupling $\delta S_{D7}/\delta \phi$ requires some care. This is because the coupling with the dilaton depends on the frame (string or Einstein frame). If one switches to  the 10-dimensional Einstein frame $G_{MN}^E=e^{-\phi/2}G_{MN}$ in (\ref{d7})
\beq
S_{D7} =-N_f T_{D7}\int d^8\bar{\xi} e^\phi \sqrt{-G^E}\left(1+\frac{(2\pi \alpha')^2}{4}e^{-\phi} {\cal F}_{ab}{\cal F}^{ab}+\cdots\right),
\eeq
one finds that the relevant coupling vanishes. However,  we actually work in the 5-dimensional Einstein frame $g^E_{MN}=e^{-4\phi/3}g_{MN}$ in the background AdS$_5$ space
\beq
S_{sugra}=\frac{1}{2\kappa^2}\int d^5x \sqrt{-g^E} \left({\mathfrak R}-12-\frac{4}{3}(\nabla \phi)^2\right).
\eeq
and in this frame the dilaton coupling is nonvanishing. To get this, write the $S^5$ part of the 10 dimensional metric in the form
\beq
d\Omega_5^2 = R^2(d\theta^2 + \sin^2 \theta d\Omega_3^2 + \cos^2 \theta d\eta^2 ).
\eeq
The $G_{zz}$ component of the brane-induced metric is then (cf. (\ref{ind}))
\beq
G_{zz}= g_{zz} + (\partial_z \theta)^2 g_{\theta\theta} =- \frac{R^2}{z^2} \left(1+\frac{z^2R^4(\partial_z \cos \theta)^2}{\sin^2 \theta}\right) =-\frac{R^2}{z^2} \left(1+\frac{z^2}{z_m^2-z^2}\right).
\eeq
This means that the 10D string frame and 5D Einstein frame are related as 
\beq
G_{zz}^S= e^{4\phi/3}g_{zz}^E + (\partial_z \theta)^2 g_{\theta\theta}^E =- \frac{e^{4\phi/3}R^2}{z^2}\left(1+\frac{e^{-4\phi/3}z^2}{z^2_m-z^2}\right) ,
\eeq
for this particular component, 
 and we find
\beq
e^{-\phi} \sqrt{-G^S} (F_{\alpha\beta}\bar{F}^{\alpha\beta})^S = e^{-\phi/3} R^3\left(\frac{R}{z}\right)^5 \left(1-\frac{z^2}{z_m^2}\right)^{3/2}\left(1+\frac{e^{-4\phi/3}z^2}{z^2_m-z^2}\right) (F_{\alpha\beta}\bar{F}^{\alpha\beta})^E.
\eeq
This leads to
\beq
\left.\partial_\phi\left(e^{-\phi}\sqrt{-G^S}(F_{\alpha\beta}\bar{F}^{\alpha\beta})^S\right)\right|_{\phi=0} &=& -R^3\left(\frac{R}{z}\right)^5 \left[\frac{1}{3} \left(1-\frac{z^2}{z_m^2}\right) + \frac{2z^2}{3z_m^2}\left(1-\frac{z^2}{z_m^2}\right) \right] (F_{\alpha\beta}\bar{F}^{\alpha\beta})^E \nn
&=& -\frac{1}{3}\sqrt{-G^E} \left(1+\frac{2z^2}{z_m^2}\right) (F_{\alpha\beta}\bar{F}^{\alpha\beta})^E.
\eeq
From this, we obtain the effective coupling
\beq
\frac{\delta S_{D7}}{\delta \phi} \propto -K_{D7} \int d\Omega_3^2 Y^l(S^3) \frac{R\phi(z)}{z} \left(1-\frac{z^2}{z_m^2}\right)\left(1+\frac{2z^2}{z_m^2}\right)\frac{\eta_{\mu\nu} \Pi^{\mu\nu}}{6}.
\eeq

\subsection{Results}
Collecting all the factors we  write 
\beq
 \langle P| \epsilon\cdot J|P'k\rangle =\bar{u}(P') \bigl( X \Pi^{\mu\nu} \Gamma_{\mu\nu} +Y \Pi^\mu_\mu \Gamma \bigr) u(P),
\label{dif}
\eeq
where (see (\ref{tt}) and (\ref{anomaly}))  
\beq
\Gamma^{\mu\nu}&=&  (A_{g}+B_{g})\gamma^{(\mu}\bar{P}^{\nu)} -\frac{\bar{P}^\mu \bar{P}^\nu}{M}B_{g}   +\frac{1}{3} \left(\frac{\Delta^\mu\Delta^\nu}{\Delta^2}-\eta^{\mu\nu}\right)\left(A_g M +\frac{\Delta^2}{4M}B_g\right), \label{ga1} \\
\Gamma&=& \frac{g}{2\beta(g)} \left((A_g-\gamma_m A_q) M + \frac{B_{g}-\gamma_mB_q}{4M}\Delta^2 -3\frac{\Delta^2}{M}(C_{g}-\gamma_m C_q) +4 (\bar{C}_{g}-\gamma_m \bar{C}_q)M \right), \label{ga}
\eeq 
and 
\beq
X&= & \lambda \frac{\kappa^2 K_{D7} }{R^2}\int d\Omega_3^2 Y^{l=0}(S^3) \int_0^{z_m} dz z^3\phi(z) \left(1-\frac{z^2}{z_m^2}\right),\\
Y&= &-\lambda \frac{\kappa^2 K_{D7}}{16R^2}\int d\Omega_3^2 Y^{l=0}(S^3) \int_0^{z_m}dz z^3\phi(z) \left(1-\frac{z^2}{z_m^2}\right) \left(1+\frac{2z^2}{z_m^2}\right) .
\eeq
$\lambda$ is a  parameter which absorbs the unknown prefactors. As already mentioned, we shall fix this by fitting the experimental data. 
Using the wavefunction (\ref{wav}) and also the formula $\int d^3\Omega Y^{l=0}(S^3)=\sqrt{2}\pi$, we find
\beq
X&=& \lambda \frac{\sqrt{2}\pi}{24} \kappa^2K_{D7} \frac{z_m^4}{R^2}, \\
Y&=&  - \lambda \frac{11\sqrt{2}\pi}{1920} \kappa^2K_{D7} \frac{z_m^4}{R^2} = -\frac{11}{80}X. \label{y}
\eeq 

The differential and total cross sections are computed from (\ref{dif}) as 
\beq
\frac{d\sigma}{dt} = \frac{\alpha_{em}}{4(W^2-M^2)^2} \frac{1}{2}\sum_{pol}\frac{1}{2}\sum_{spin}|\langle P| \epsilon\cdot J|P'k\rangle    |^2    , \qquad  \sigma_{tot}= \int_{t_{min}}^{t_{max}}dt \frac{d\sigma}{dt}.  \label{cro2}
\eeq
The first sum is over the photon and $J/\psi$ polarizations. This  can be done according to the formula
\beq
\sum_{s=1,2} \epsilon^\mu_{s}\epsilon^{*\nu}_{s} \to - \eta^{\mu\nu} \quad  \sum_{s'=1,2,3} \xi^\mu_{s'}\xi^{*\nu}_{s'} = -\eta^{\mu\nu} + \frac{k^\mu k^\nu}{M_\psi^2}.
\eeq 
The second sum is over the initial and final proton spins.  
Defining $\Pi^{\mu\nu}\equiv \Pi^{\mu\nu}_{\alpha\beta} \epsilon^\alpha \xi^\beta$, we get  
\beq
I_s &\equiv& \sum_{pol}\sum_{spin}|\langle P| \epsilon\cdot J|P'k\rangle    |^2 \nn
 &=& {\rm Tr} \Bigl[ \bigl( X\Pi^{\mu\nu}_{\alpha\beta} \Gamma_{\mu\nu} +Y (\Pi^\mu_\mu)_{\alpha\beta} \Gamma \bigr) (\Slash P+M)  \bigl(   X\Pi^{\mu'\nu',\alpha\beta} \Gamma_{\mu'\nu'} +Y (\Pi^{\mu'}_{\mu'})^{\alpha\beta} \Gamma \bigr)  (\Slash P'+M)\Bigr] \nn 
&& \quad -\frac{k^\beta k^\gamma}{M_\psi^2}{\rm Tr} \Bigl[  \bigl( X\Pi^{\mu\nu}_{\alpha\beta} \Gamma_{\mu\nu} +Y (\Pi^\mu_\mu)_{\alpha\beta} \Gamma \bigr) (\Slash P+M)  \bigl(   X\Pi^{\mu'\nu',\alpha}_{\ \ \ \ \ \ \gamma} \Gamma_{\mu'\nu'} +Y (\Pi^{\mu'}_{\mu'})^{\alpha}_{\ \gamma} \Gamma \bigr)  (\Slash P'+M) \Bigr].  \label{fin}
\eeq
We computed (\ref{fin}) using {\sf Feyncalc} and expressed the result in terms of $W^2=M^2+2P\cdot q$ and  $t=M_\psi^2-2q\cdot k$. (Note that $P'=P+q-k$ and  $P\cdot k= \frac{W^2+t-M^2}{2}$.).   The full analytical expression turns out to be too lengthy to be reproduced here, but the following points are worth noting:   (i) Formally, the result can be Laurent expanded in $t$ as 
\beq
I_s=\frac{A_g^2M^4M_\psi^8 X^2}{9t^2}  &+& \frac{A_g M^2 M_\psi^4 X}{36t}\Bigl[A_g X \left(24M^4+16M^2(M_\psi^2-3W^2)+5M_\psi^4-24M_\psi^2 W^2+24W^4\right) \nn 
&& \qquad +2B_g M_\psi^4 X + 96MM_\psi^2 Y\Gamma\Bigr]+ {\cal O}(t^0). \label{tr}
\eeq
One immediately recognizes  an apparent  singularity $1/t^2$ and might worry that such a rapid behavior of $d\sigma/dt$ at small-$t$ is at odds with the experimental data. However, this is totally innocuous. In practice $|t|$ cannot be smaller than the value determined from  (\ref{min}), and when $t=t_{min}$, the `singular' terms in (\ref{tr}) are numerically comparable, or even smaller than the other `nonsingular' terms.\footnote{One might wonder why  poles in $1/t$ appear  although there is no divergence in the limit $\Delta\to 0$ at the amplitude level (\ref{ga1}). The answer is that the limit $\Delta\to 0$ has to be taken together with the (unphysical) limit $M_\psi\to 0$ in order to be kinematically consistent. Note that poles in $1/t$ are proportional to  $M_\psi$.  }  (ii) If one expands $I_s$ in  $W^2$, one finds that the highest power is $W^8$. When combined with the prefactor  $1/W^4$ in (\ref{cro2}), this gives a very strong energy dependence $d\sigma/dt \sim  s^2$ at large $s=W^2$. This is an artifact of the graviton exchange which is a spin-2 particle.  We are not concerned about this asymptotic behavior, since our focus is near the threshold region $W\gtrsim W_{th}$ where the graviton and dilaton contributions are comparable. 

For a numerical evaluation, we use $M=0.94$ GeV, $M_\psi=3.1$ GeV and assume the dipole form for the gravitational form factors\footnote{While the dipole form for $A_{q,g}(t)$ is commonly used (e.g., \cite{Frankfurt:2002ka}), we are not aware of any literature which discusses the $t$-dependence of  $\bar{C}_{q,g}$. Eq.~(\ref{thus}) is just an assumption which should be used with care at large-$t$.} 
\beq
A_{q,g}(t) = \frac{A_{q,g}(0)}{(1-t/\Lambda^2)^2}, \quad \bar{C}_g(t)= \frac{\frac{1-b+\gamma_m}{1+\gamma_m}-A_g(0)}{4(1-t/\Lambda^2)^2} =-\bar{C}_q(t), \label{thus}
\eeq
with $\Lambda^2 = 0.71$ GeV$^2$. We fix $A_g(0)=0.43=1-A_q(0)$ \cite{Yang:2017erf} and vary the parameter $1\ge b \ge 0$. 
As seen in (\ref{anomal}), when $b=0$ the trace anomaly contributes maximally to the proton mass, whereas $b=1$ corresponds to vanishing anomaly contribution.  A recent model calculation has found a rather small value for $\bar{C}_g(0)=-\bar{C}_q(0) \sim {\cal O}(10^{-2})$ \cite{Polyakov:2018exb}. As for   $B_g(t)$, we simply neglect it  following indications \cite{Hagler:2007xi,Selyugin:2009ic,Alexandrou:2017oeh} that $B_q(0)=-B_g(0)$ happens to be numerically very small. Unfortunately, almost nothing is known about $C_g(t)$, or the gluon D-term $D_g(t)=4C_g(t)$. We employ a simple model inspired by the asymptotic behavior and the quark counting rule \cite{Tanaka:2018wea}\footnote{Incidentally, in the present model $I_s$ behaves as $t^{5}C_g^2(t)$ at large $t$, so the strong falloff $C_g\sim 1/t^3$  as predicted by the counting rule \cite{Tanaka:2018wea}   is needed to ensure that $d\sigma/dt$ is a decreasing function of $t$ at large-$t$. On the other hand, in the small-$t$ region, different powers of $t$, or even the exponential form $C_g(t)\sim e^{bt}$ are indistinguishable in practice \cite{Frankfurt:2002ka}.  }  ($n_f=3$ here)
\beq
C_g(t) = \frac{16}{3n_f}C_q(t)   = \frac{16}{3n_f} \frac{-0.4}{(1-t/\Lambda^2)^{3}}, \label{ba}
\eeq
 where the value $4C_q(0)=D_q(0)\approx -1.6$ is taken from \cite{Pasquini:2014vua}. As one might expect from the explicit $\Delta^2=t$ factor in the coefficient (\ref{ga}), the effect of the $C_g$-term is minor in the small-$t$ region, $\sqrt{|t|}< 1$ GeV, whereas it becomes significant at large $\sqrt{|t|}> 1$ GeV. Very close to the threshold, $\sqrt{|t_{min}|} \sim 1$ GeV (see Fig.~\ref{fig1}), so the uncertainties in $C_g$ should not be underestimated. Finally, we assume fixed coupling $g=2$ ($\alpha_s\approx 0.32$) with $N_c=n_f=3$ in the 1-loop beta function so that the prefactor in (\ref{ga}) becomes $\frac{g}{2\beta}\approx -2.2$. The mass anomalous dimension evaluated at the same order is $\gamma_m = \frac{2\alpha_s}{\pi} \approx 0.2$.
Of course, all these form factors should be modeled in more sophisticated manners \cite{Goeke:2007fp,Wakamatsu:2007uc,Selyugin:2009ic,Abidin:2009hr,Chakrabarti:2015lba}. We leave this to future work. 

\begin{figure}[h]
 \includegraphics[height=45mm]{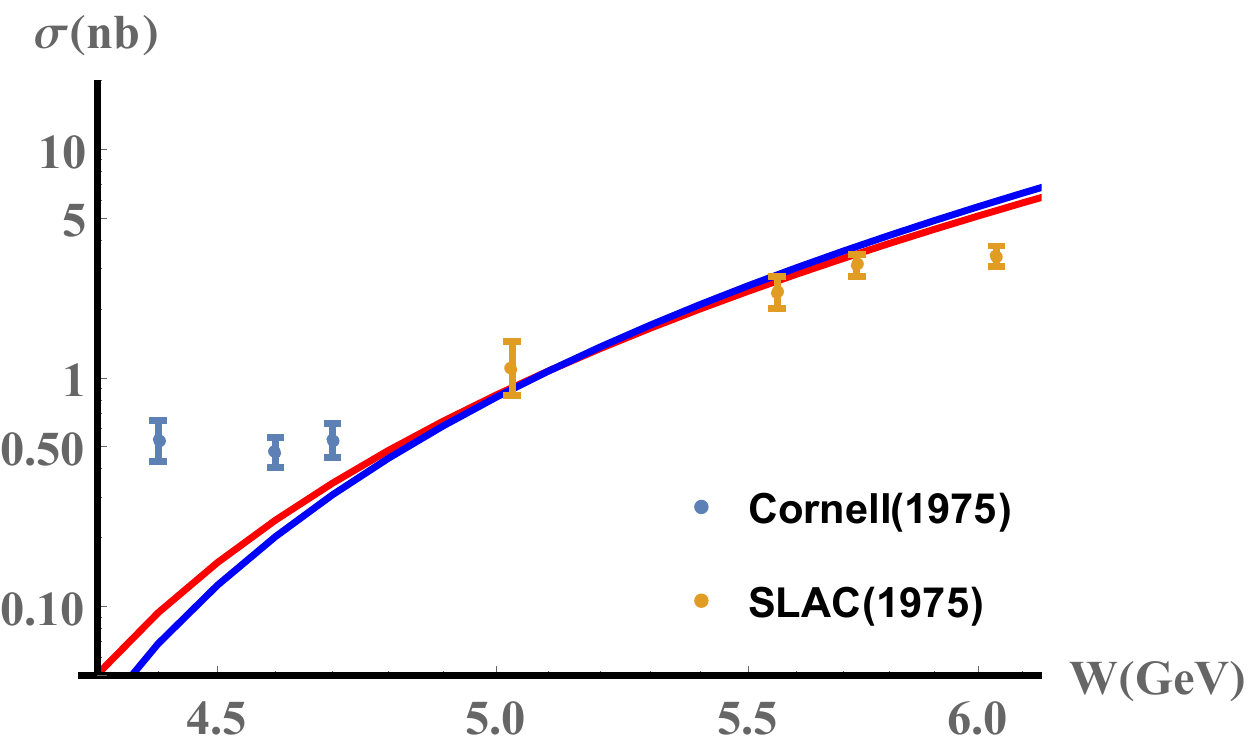}
  \includegraphics[height=45mm]{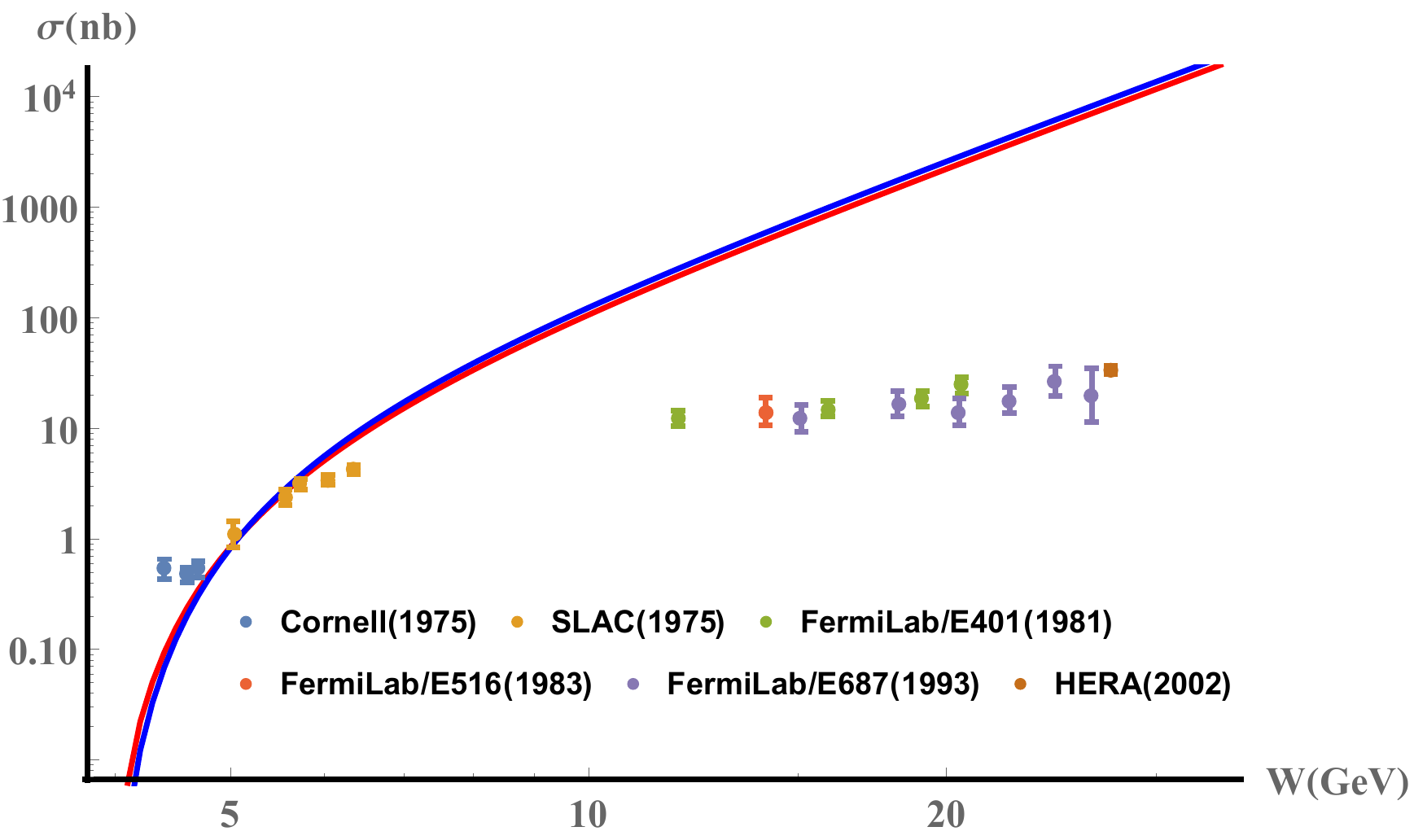}
\caption{ The total cross section in units of nb as a function of $W$. The upper red curve: maximal anomaly contribution. Lower blue curve: zero anomaly contribution.
} 
\label{fig2}
\end{figure}

\begin{figure}[h]
 \includegraphics[height=45mm]{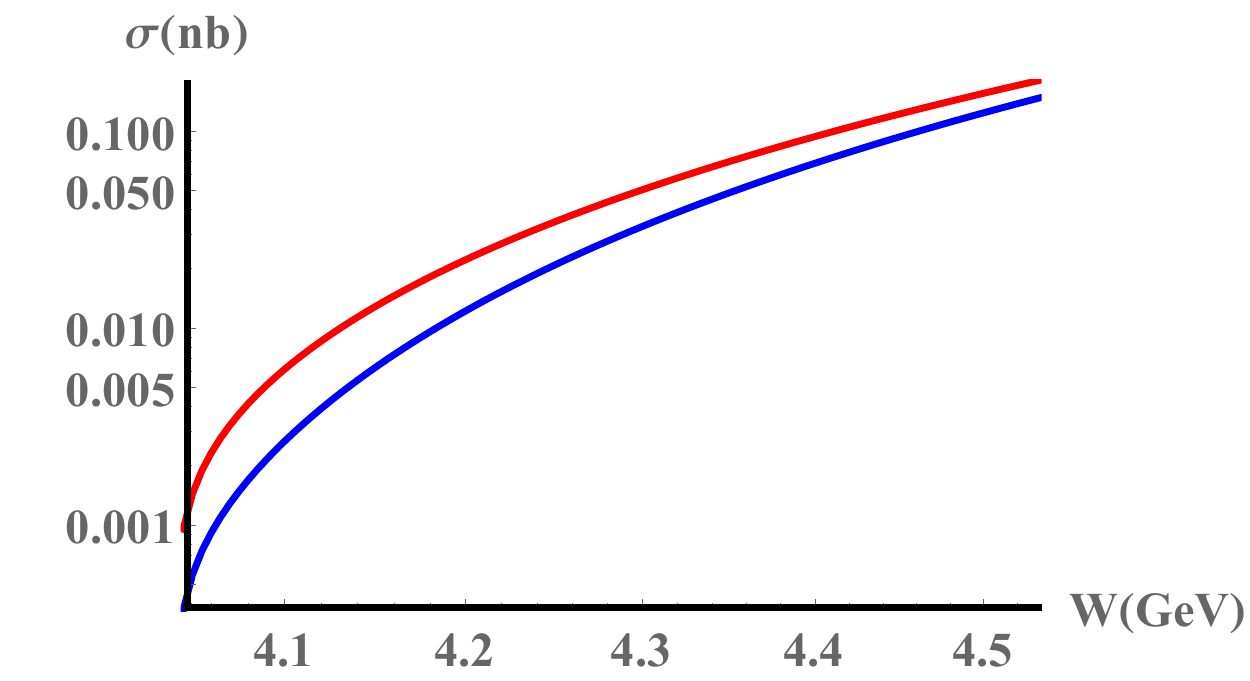}
\caption{ The total cross section very close to the threshold.
} 
\label{fig3}
\end{figure}

We first plot in Fig.~\ref{fig2} the total cross section $\sigma_{tot}$ as a function of $W$ and compare with the experimental data from Cornell \cite{Gittelman:1975ix}, SLAC \cite{Camerini:1975cy}, Fermilab \cite{Binkley:1981kv} and HERA \cite{Chekanov:2002xi} as  summarized  in \cite{Gryniuk:2016mpk}. The overall normalization factor has been fixed by  performing a $\chi^2$ fit of the low energy $(W\le 6$ GeV) data points. 
The upper red curve corresponds to $b=0$ (maximal anomaly) and the lower blue curve corresponds to $b=1$ (zero anomaly).  The effect of the trace anomaly is  visible only near the threshold $W\lesssim 5$ GeV.  As expected, the graviton exchange gives a too strong rise of the cross section $\sigma_{tot}\sim W^4=s^2$ in the high energy region where the experimental data show a much milder growth. This is due to the different nature of `Pomeron' between  QCD and gravity theories, and there are many attempts in the literature to correct for this differece.  Our focus, instead, is on the low energy regime where the $W$-dependence in the SLAC region is roughly reproduced. However, we have difficulty in fitting the Cornell data points which are almost flat in $W$. It should be kept in mind that these old data points suffer from low statistics and the lack of exclusivity, and should be revised in future experiments \cite{Joosten:2018gyo}. We hope to redo our fit when new data become available. 
 Fig.~\ref{fig3} shows $\sigma_{tot}$ very close to the threshold $W\lesssim 4.5$ GeV. In this regime, the trace anomaly can enhance the cross section by a factor of 2 or more. 

\begin{figure}[h]
 \includegraphics[height=45mm]{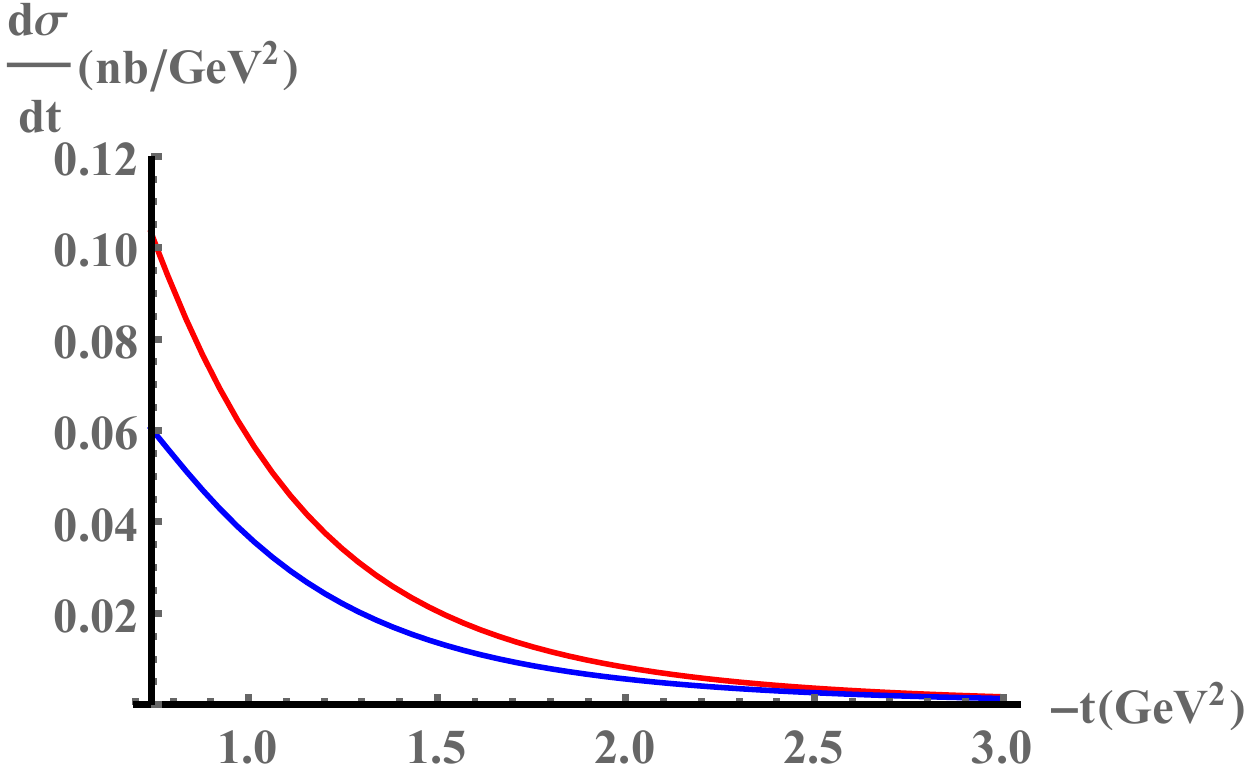} 
 \includegraphics[height=45mm]{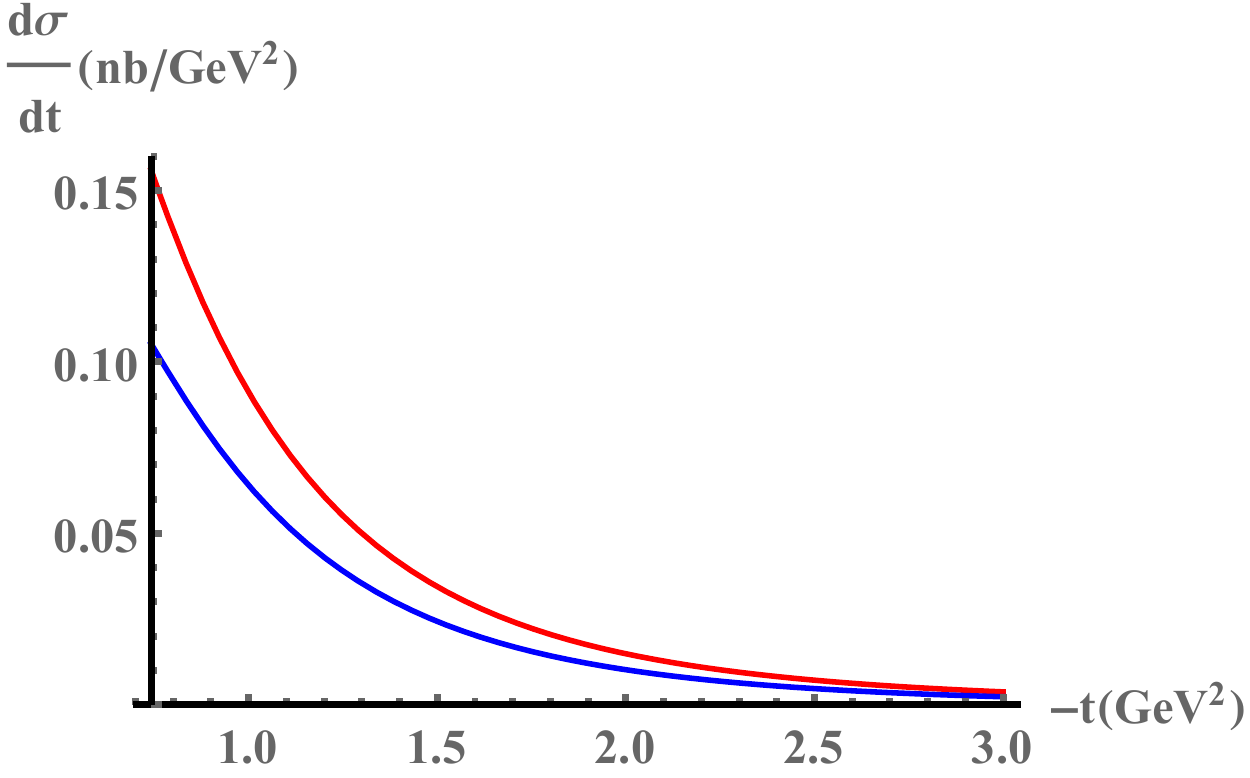} 
\caption{ The differential cross section at $W=4.3$ GeV. $C_g\neq 0$ (left) and $C_g=0$ (right). 
} 
\label{fig4}
\end{figure}

Next we plot $d\sigma/dt$ as a function of $t$ at $W=4.3$ GeV (Fig.~\ref{fig4}). On the right panel, we artificially set $C_g(t)=0$ to see the impact of this poorly constrained function. We clearly see the effect of the trace anomaly on the shape of the distribution $d\sigma/dt$. With the anomaly (upper curve), $d\sigma/dt$ is enhanced at small $t$, and it falls off more rapidity with $|t|$. This tendency is more pronounced as one decreases $W$ and approaches the threshold. Note however, that closer to the threshold the uncertainty due to the $C_g(t)$ term also becomes larger. Although this mostly affects the overall normalization rather than the shape, more serious models  of $C_g(t)$ in this region are certainly welcome.

\begin{figure}[h]
 \includegraphics[height=51mm]{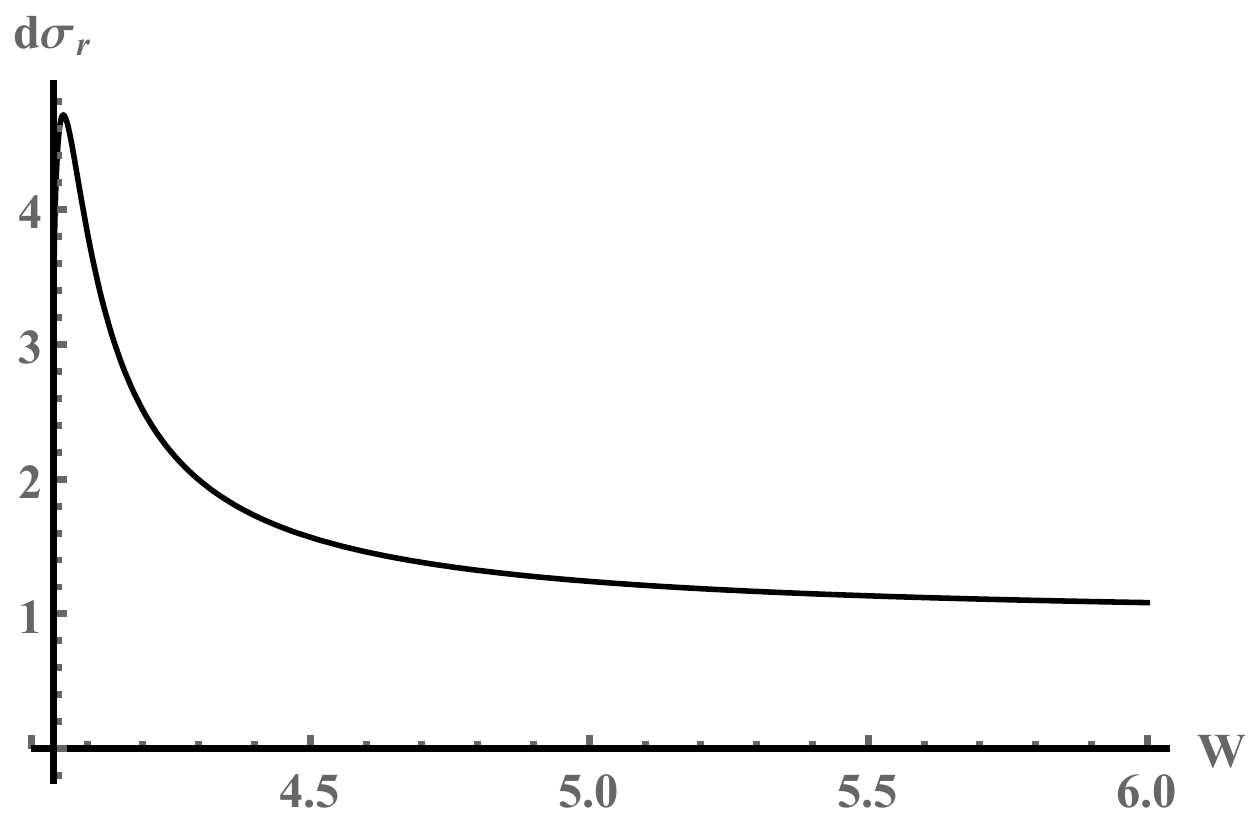} 
\caption{ The ratio (\ref{rat}) evaluated at $t=t_{min}$. 
} 
\label{fig5}
\end{figure}

Finally, in Fig.~\ref{fig5} we plot the following ratio
\beq
d\sigma_r \equiv \frac{\left(\frac{d\sigma}{dt}\right)_{b=0}}{ \left(\frac{d\sigma}{dt}\right)_{b=1}}, \label{rat}
\eeq
evaluated at $t=t_{min}$ as a function of $W$. This plot shows that the effect of the trace anomaly is largest when $W\approx 4.06$ GeV where it enhances the peak value of $d\sigma/dt$ by a factor of about 4. In order to explore the peak region in Fig.~\ref{fig5}, one should tune the collision energy $W$ to be less than $4.5$ GeV (or $E_\gamma\lesssim 10$ GeV in the proton rest frame).  At larger energies, the ratio flattens but stays larger than unity.

\section{Conclusions}

In this paper we have undertaken the first study of the detailed relation between the $J/\psi$ production cross section and the QCD trace anomaly from holography. The key observation is that the trace anomaly enters the cross section via the dilaton exchange, and we have 
shown how it is related to the matrix element of $T^{\mu\nu}$. Our findings carry important messages to the experimentalists who are planning to measure this process. Firstly, the center of mass energy $W$ should be  $W\lesssim 4.5$ GeV, in order to clearly see the effect of the trace anomaly. At higher energies the dilaton contribution is overwhelmed by the graviton contribution. Once the energy is chosen in this regime, the shape of the differential cross section contains information about the trace anomaly. Of course, in reality one cannot turn on and off the anomaly contribution to see the difference. But at least one can compare with model predictions without the trace anomaly  to see if there are noticeable differences, especially in the peak value of $d\sigma/dt$ at $t=t_{min}$. One caveat is that if $W$ is too close to the threshold, uncertainties in the $C_g$ term (or the gluon D-term) become large. More theoretical work is needed to constrain this form factor. 

Our study also shows that the $t$-dependence is not the exponential form $d\sigma/dt \sim e^{bt}$ as is often assumed. It is not purely that of the square of some form factors  \cite{Frankfurt:2002ka}, either.   The cross section does contain the square of  various gravitational form factors, but they are multiplied by  complicated (but rational) functions of $t$. Moreover, our result for $d\sigma/dt$ cannot be naively extrapolated to $t\to 0$ because of the presence of the $1/t^2$ term. Nevertheless, if one knows the  $t$-dependence of various nucleon form factors, one can extract the trace anomaly parameter $b$ from the experimental results at finite-$t$.

There are many directions for future studies. We have used the simplest setup, namely, D7 branes embedded in an asymptotically AdS$_5$ space, or the `D3/D7 model.' It would be very interesting to study the present process in  more realistic AdS/QCD models.  Also, more precise parameterizations  of the form factors $A_g,B_g,...$ are certainly important to confront the experimental data. These gluonic form factors are quite difficult to access, but there has been steady progress in the QCD community toward this goal. Finally, it is important, but quite challenging to include the stringy effects beyond the supergravity approximation. Once the stringy effects are included, we expect that the amplitude becomes complex-valued \cite{Brower:2006ea,Hatta:2007he}. In the case of Deeply Virtual Compton Scattering (DVCS), an extensive discussion can be found in  \cite{Nishio:2012qh,Nishio:2014eua,Nishio:2014rya}. The present process could also be studied in such a framework.

\section*{Acknowledgements}
We are grateful to Maxim Polyakov for pointing out a mistake in an earlier version of this paper. 
We also thank Oleksii Gryniuk, Dima Kharzeev, Elias Kiritsis,  Zein-Eddine Meziani, Barbara Pasquini, Jianwei Qiu, Shigeki Sugimoto and Kazuhiro Tanaka for discussion and correspondence. Y.~H. is supported by the U.S. Department of Energy, Office of Science, Office of Nuclear Physics, under Contracts No. de-sc0012704. D.~Y. is supported by the RIKEN Foreign Postdoctoral Researcher program.

\appendix 

\section{Propagators in AdS}
\label{a1}

Consider the massless scalar (dilaton) action in the AdS$_5$ background 
\beq
S_\phi=\frac{c}{2\kappa^2} \int d^5x \sqrt{-g} \frac{1}{2}(\nabla \phi)^2,
\eeq
where $c$ is some constant and $2\kappa^2= \frac{8\pi^2R^3}{N_c^2}$.  In the five-dimensional Einstein frame, $c=\frac{8}{3}$. 
The bulk-to-bulk  propagator is 
\beq
D(xz;x'z') = \langle \phi(xz)\phi(x'z')\rangle = \frac{2\kappa^2i}{cR^3} \frac{3}{2\pi^2} \frac{1}{(2u)^4} F\left(4,\frac{5}{2},5;-\frac{2}{u}\right),
\eeq
 where 
\beq
u=\frac{(z-z')^2 -(x-x')^2}{2zz'},
\eeq
is the chordal distance in AdS$_5$. Taking the limit $z\to 0$, we find
\beq
D(x,z\to 0,x'z') \approx \frac{2\kappa^2 i}{cR^3} \frac{3}{2\pi^2} \left( \frac{zz'}{z'^2-(x-x')^2+i\epsilon}\right)^4 . \label{bul}
\eeq
Now consider the gauge theory matrix element $\langle P|\frac{1}{4}F^{\mu\nu}_a F^{a}_{\mu\nu}(x)|P'\rangle$.  
The insertion of the operator $F^2$ at the boundary point $x$ excites a dilaton field excitation in the bulk 
\beq
\phi(x'z')=   \frac{6i}{\pi^2} \left( \frac{z'}{z'^2-(x-x')^2+i\epsilon}\right)^4 . \label{nor}
\eeq
(This is normalized as $\phi(x'z'\to 0) = \delta^{(4)}(x-x')$.) 
The point is that (\ref{bul}) and (\ref{nor}) are simply proportional to each other. 
Thanks to this, we may approximate the proton side of the amplitude $\int d^4x'dz' D(xz,x'z') \delta S_B/\delta \phi$ by $\langle P|\frac{1}{4}F^2|P'\rangle$  after taking into account the difference in the prefactor $\sim z^4$. 

Specifically, the expectation value of $F^2$ is given by the variation of the on-shell dilaton action in the presence of a  source  (proton) \cite{Danielsson:1998wt}\footnote{
Since the relative sign between the graviton and dilaton exchanges is important, let us quickly check the sign in (\ref{on}). 
The D3 brane action is 
\beq
S_{D3}=-T_{D3} \int d^4x e^{-\phi} {\rm Tr} \sqrt{-{\rm det}(G+2\pi \alpha'F)} &\sim&  -\int d^4x e^{-\phi} \sqrt{-G} \frac{2}{4g^2}{\rm Tr}F^{\mu\nu}F_{\mu\nu} \nn  &=&  -\int d^4x e^{-\phi} \sqrt{-G} \frac{1}{4}F^{\mu\nu}_a F^a_{\mu\nu},
\eeq
where $T_{D3}=\frac{1}{8\pi^3\alpha'^2 g_s }$ and $4\pi g_s = g^2$. $F$ here denotes the SU($N_c$) gauge field. In the last line we rescaled $F/g\to F$ which is the standard normalization in QCD used in earlier sections. We thus find 
\beq
 \frac{1}{4}F^{\mu\nu}_a F^a_{\mu\nu}  = \frac{\delta S}{\delta \phi}.
\eeq
} 
\beq
\langle P|\frac{1}{4}F^{\mu\nu}_a F^{a}_{\mu\nu} |P' \rangle =  \frac{\delta S_\phi}{\delta \phi(z=0)} =\left. \frac{cR^3}{2\kappa^2}\frac{1}{z^3} \partial_z \phi \right|_{z=0}. \label{on}
\eeq
The bulk field $\phi$ is determined by solving the equation of motion 
\beq
-\frac{c}{2\kappa^2}\sqrt{-g}\nabla^2 \phi + \frac{\delta S_{B}}{\delta \phi} = 0,
\eeq
with the solution
\beq
\phi (xz)= \int d^4x'dz'  iD(xz,x'z') \frac{\delta S_B}{\delta \phi}.
\eeq
Substituting this into (\ref{on}) and noting that
\beq
\frac{1}{z^3}\partial_z D(xz,x'z') \approx \frac{4}{z^4}D(xz,x'z'),
\eeq
 because $D(z,z')\propto z^4$  as $z\to 0$, we find the following correspondence
\beq
\langle P| \frac{1}{4}F^2 |P'\rangle  \approx \frac{cR^3}{2\kappa^2}\frac{4}{z^4}\int d^4x'dz'iD(xz,x'z')\frac{\delta S_B}{\delta \phi}.
\eeq


Similarly, the bulk-to-bulk graviton propagator can be approximately replaced by the expectation value of $T_{\mu\nu}$.  
The metric perturbation  due to the source term $S_B$ is 
\beq
\delta g_{MN}(xz) = \int d^4x'dz' \, iG_{MN,M'N'} \frac{\delta S_B}{\delta g_{M'N'}}. 
\eeq
From this one can read off the matrix element of the energy momentum tensor via the `holographic renormalization' \cite{deHaro:2000vlm}
\beq
\langle  T_{\mu\nu}\rangle = -\frac{R^3}{2\kappa^2} \lim_{z\to 0} \frac{4}{z^4} \left(\frac{z^2}{R^2}\delta g_{\mu\nu}\right). 
\eeq
 (The minus sign is because we use the mostly minus metric $\eta^{\mu\nu}=(1,-1,-1,-1)$.)

\end{document}